# A Method to Decipher "Genome" from Interatomic Cohesion in the Exploration for a "Central Dogma" Replacement in Material Science

*Xinxu Zhang[1], Jiahao Wei[1], Hui Jia[1], Jiamin Liu[1], Guo Li[1], Ling Liu[3], Yulong Wu[1], Changlong Liu[1], Xiao-Dong Zhang[1,2], Yonghui Li[1*]*

1. Department of Physics and Tianjin Key Laboratory of Low Dimensional Materials Physics and Preparing Technology, School of Sciences, Tianjin University, Tianjin 300350, China
2. Tianjin Key Laboratory of Brain Science and Neural Engineering, Academy of Medical Engineering and Translational Medicine, Tianjin University, Tianjin 300072, China

**ABSTRACT:** In the ball-stick model, interatomic cohesions are considered "sticks". But enormous details and features of the "sticks" are usually oversimplified as indexed quantities or equivocated as geometry characteristics. These indexed quantities or geometry characteristics not only limit the explanatory capability to a few chemical/physical aspects but also eliminate generativity for expected resemblance. And these limitations can be related to the information loss during the conversion. Herein, inspired by the central dogma, a framework is introduced to compact interatomic cohesions into a detailed residue-by-residue "genome" with matched encoding/decoding tools. The framework fuses the quantum mechanical aspects, auto feature extraction, nanostructures and/or simulations, and generative models. As a proof of concept, the realization introduced in this work adopted bosonic/fermionic features, an autoencoder with image recognition processes, Density Functional Theory simulations, and a thiolate-protected gold nanocluster dataset. After repetitive modeling, validating, and analysis based on 26,528 simulated interatomic images, the interatomic cohesion can be almost losslessly encoded into an 8-value-genome, and the genome encoder-decoder pair is also obtained. The model is then automatically extended into a generative model which converts any arbitrary 8-value-genome to a bond image. With a detailed exploration of the space span by the 8-value-genome, a variety of details are observed including bond polarization, hybridization, intrusion of other atoms, alignments, crystal orientation, atomic motions, and more. In the 8-value-genome representation, molecules and solids are unified in a briefer way than geometric quantities. Therefore, the interatomic genome encoder and decoder are not just a collector of superficial bond statistics but a coherent model that "understands" human knowledge of chemical bonds. In the last of this work, a roadmap is plotted by summarizing and correlating the similarities of 8-value-genomes. 3 sets of applications including a set of chemisorption, a set of molecular dynamical analysis, and a set of ultrafast processes are provided. Though for demonstration purposes and limitation to the dataset, this work only focuses on a narrow but challenging mesoscopic family of nanoclusters, the framework itself can be broadcasted to a variety of scenarios and extended into the whole periodic table for the next generation of quantitative description of interatomic cohesion.

## I. Introduction

When we consider condensed matters as building blocks of balls with "connecting pieces" (i.e., interatomic cohesion), these puzzling connectors should be sticks with mystic electrostatic interactions and quantum effects. With the development of science, these connectors, acknowledged as chemical bonds[1] including ionic, covalent, metallic, etc. are defined based on a coarse interpretation of electron structures. As the comprehension progressively intensified, the Valence Shell Electron Pair Repulsion (VSEPR) theory[2,3] was introduced and aimed to claim the repelling (acknowledged as Fermionic side years after) of the interatomic cohesion. With the development of quantum theories, the gap between the interatomic cohesion and electron wavefunctions is partially bridged by the introduction of the Electron Localization Function (ELF)[4–10], but the massive information contained prevents us from seeing these "connecting pieces" clearly. Despite continuous efforts to refine and quantification over time[11–13], building matters with atoms and connectors is still a dream of humans.

The nature of the interatomic cohesion with its topology is deeply rooted in the fermionic facts of electrons with massive information contained. With the development of machine learning, such critical aspect is further pushed into the intersection of physics, chemistry, and data science. As shall be explained, we may not need a concept as the chemical bond to aid the dream of matter construction but using a "genome of interatomic cohesion" (interatomic genome) instead can fulfill the task. Though humans are less sensitive to the mild change in interatomic interactions than that in face recognition, the interatomic genome can provide more explanative details.

In this work, we push the machine-level interatomic cohesion description to a new limit with both abstractive and generative aspects. On the theoretical side, it is a framework to convert any arbitrary bond to an interatomic genome as its numerical representation which is featured by precision, comprehensiveness, and continuity. The shift of the interatomic genome traces the change in the topology of interatomic cohesion precisely.

In practice, the interatomic genome should outperform representations of images or spatial distributions in the quantification of interatomic cohesions. As we shall demonstrate, our realization of the framework survived the test of extremely challenging mesoscopic materials that have no global patterns. To demonstrate, the well-trained generator produces interatomic cohesions in thiolate-protected gold nanoclusters[14–16] (TP-AuNCs) as images **(Figure 1a)** mimicked or "machine imaged". On the level of visual Turing tests, since humans cannot tell the differences, the representation and algorithm prove their internal comprehensibility. On the interatomic genome side, the well-trained encoder extracts the interatomic genome precisely from the interatomic cohesion **(Figure 1b)**. As we shall extensively explain constructing interatomic genomes is a feasible and robust foundation for further investigations.

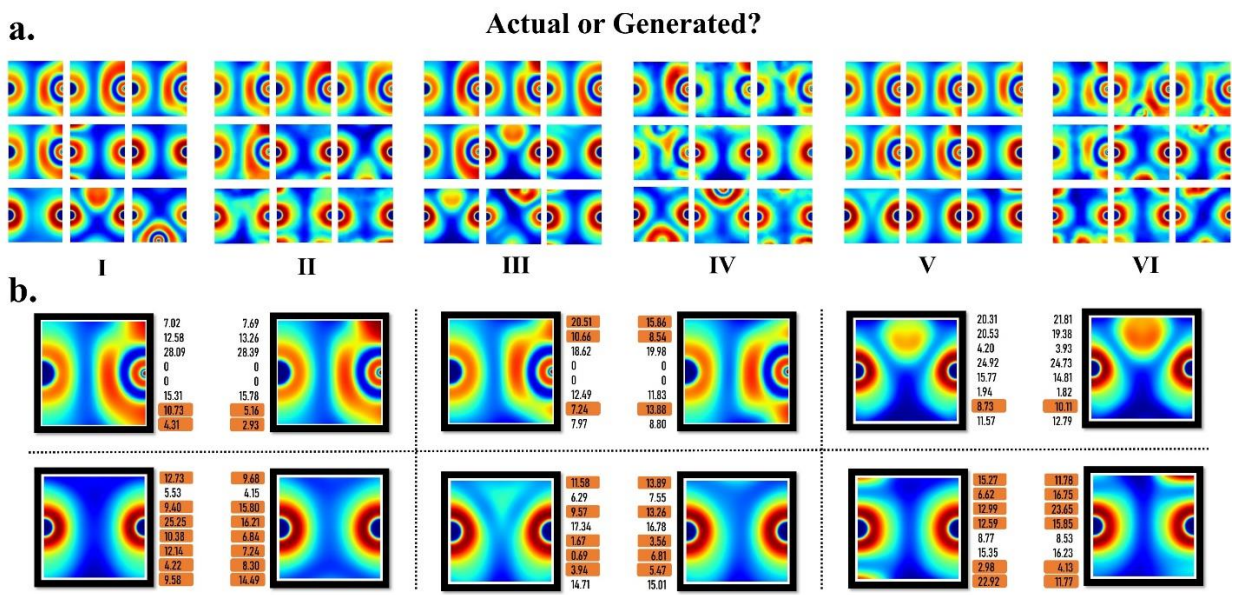

**Figure 1.** Visually Distinguishability vs. Interatomic Genome **(a):** The ELF images of S-Au and Au-Au interactions in various TP-AuNCs. QUIZ: Can you tell the generated ones from quantum calculated ones? Visually telling generated images and images from quantum mechanics is challenging in some cases. (ANSWER: Quantum calculated=I, III, and V; generated=II, IV, and VI). **(b):** Some visually similar chemical bond images of S-Au and Au-Au interactions reflect significant differences in their interatomic genome (see genes in red background) which covers different natures as shall be explained later in this work.

### II. Indexing Algorithms in a "Trilemma"

Inspired by data mining processes and general rules (see SI for more details), we can summarize a Trilemma among 3 key objectives that a scientist must confront when extracting critical features in chemical bonds as massive information contained quantities (including electron densities, wave functions, various energies, etc.). These three objective angles, which cannot all be fully attained simultaneously, are:

a) Descriptiveness which quantitatively tells specific properties of materials.
b) Generality which describes the coverage and common properties across various types of materials.
c) Model briefness which refers to the complexity in processing data in chemical bonds. (See SI for the definition.)

In the scope of the "Trilemma", numerous existing methods for assessing/indexing/characterizing chemical bonds can be reviewed. Since the introduction of quantum mechanical perspectives into the description of interatomic interactions by Heitler and London[17], a profusion of chemical bond theories and models have emerged rooted in the electron wave functions and their derived physical quantities. From the perspective of the "chemical bond Trilemma", contemporary chemical bond analysis methods must strike a balance among the three aspects.

First, let's take the Natural Bond Orbital (NBO) approach[18–20], which employs classical Lewis model constructs for chemical bond analysis with its following subsequent theories such as Natural Resonance Theory (NRT)[21–23] and Adaptive Natural Density Partitioning (AdNDP)[24] as a whole example. The NBO analysis builds upon performing partial diagonalization of the electron density matrix and employing algorithms to identify valence shell orbitals, combined with related descriptors. Its computational complexity allows for a detailed description of bonding interactions in alkali metals and alkaline earth metals[25]. However, it exhibits inherent theoretical limitations when dealing with transition metals and certain main group elements[26–28]. Thus, we consider that NBO strikes a careful balance among descriptiveness (classical Lewis model, polarity, and hybridization), generality (mainly covers alkali and alkaline earth metals), and model briefness (partial diagonalization of the electron density matrix). Its "fair-poor" briefness (in the sense of complicity) leads to a relatively good descriptiveness and generality.

In comparison, the quantum theory of Atoms in Molecules (AIM)[29,30] proposed by Richard Bader likely represents one of the most physically grounded approaches in the analysis of chemical bonding. This method derives its strength from modeling and analyzing the electron density ($\rho$) at Bond Critical Points (BCPs), leading to a set of widely adopted descriptors for chemical bonding. Cremer and Kraka utilize the energy density H(**r**) at BCPs as a criterion for assessing chemical bond characteristics[31]. Nonetheless, the numerous indicators defined by this method are often perceived as somewhat lenient, which can lead to inaccuracies in predicting actual chemical reactions. So, we value AIM as an algorithm with good briefness (easily formulated and calculated), and good generality (various applications across the periodic table) but poor descriptiveness.

Unlike charge partitioning techniques (NBO and AIM), an energy partition algorithm, Energy Decomposition Analysis-Natural Orbitals for Chemical Valence (EDA-NOCV) method[32,33] focuses on energy decomposition by breaking down the total energy of a molecule into several interactional contributions between atoms, including orbital interactions, electrostatic interactions, exchange effects, and more[32]. This makes the EDA-NOCV model not only capable of quantitatively representing the strength of orbital interactions but also visually illustrating the related charge transfer processes. There are currently numerous reported cases that highlight the significant role the EDA-NOCV method has played in analyzing interatomic interactions within a diverse array of main-group compounds[34,35] and transition metal complexes[36,37]. As indicated by various works the EDA-NOCV could be an excellent method, limited to a specific range of applications (generality: majorly targeted for charge transfer systems). The briefness of EDA-NOCV is comparable to NBO in the level of "fair-poor". The descriptiveness is very convincing and meaningful.

Moreover, for the description of metallic bonds, Hammer and Noskov's d-band center model[38,39] is currently widely and successfully employed in the interpretation and prediction of catalytic activities on various transition metal surfaces. This theory posits that when an adsorbate approaches the surface of a metal catalyst, its electron cloud interacts with the metal's d-orbitals. The separation between the d-band center (the average energy level of the d-band of orbitals) and the Fermi energy is positively related to the adsorption strength. This model is algorithmically simple (great briefness) and offers a highly descriptive account of the chemical adsorption capabilities of transition metal surfaces for small molecules[40] (great descriptiveness). However, it does exhibit severe limitations in terms of generality, as it is primarily applicable only to transition metal materials[41,42].

With the advancement of machine learning algorithms and big data, researchers are now able to handle and apply more complex algorithmic models. And it is the time for people to pursue indicators with both great descriptiveness and generality. Thus, characterizing models must be polymorphic and complex and should be evolved in a data-driven way with 1) a sensitive and precise representer (e.g. extracted genomes in this work), 2) a large enough coverage in various bonds, and 3) a training-validation criterion. As shall be demonstrated in the following paragraphs, the genomes that are obtained from a machine learning-based model provide a solid foundation for the next generation of characterizing representers as well as facilitate further generative models. As shall be demonstrated, the machine learning model possesses an ideal generality. The extracted genome can be applied seamlessly to property prediction rules for potential excellent descriptiveness. The price paid for pursuing generality and descriptiveness to their extremes is the worst briefness including a massive amount of chemical bond data and tough model training and optimizing. Though there are deep learning models for analyzing bonds for a specific type of materials[43,44], there is no extensive discussion for chemical bond-oriented models.

### III. An Interatomic Genome Decipher
### A. The framework

The interatomic genome framework focus on small topology changes in every interatomic cohesion. It fused the quantum mechanical aspects, auto feature extraction, nanostructures and/or simulations, and generative models.

On the quantum mechanical side, behaviors of electrons can be split into bosonic and fermionic patterns as the two sides of a coin. Due to the Pauli exclusion principle, electrons are spatially exclusive to each other, resulting in "exchange-correlation holes[45]" (in the language of the single particle approximation) to prevent trespass of other electrons. While out of these exclusive holes, the behavior of electrons is similar to bosons. Based on the bosonic-fermionic split approach, ELF has become a convincing strategy. Unfortunately, human-comprehensive formulation which makes ELF famous becomes the bottleneck for data-oriented applications. Since the fermionic pattern in ELF is scaled to a very small range, it may lead to ignorance of the fermionic patterns. As explained, focusing on the bosonic/fermionic patterns could be out best bet, using other aspects as representations are possible (see SI section b). This framework is extendable to further theoretical/experimental interatomic description.

Considering the feature extraction component, the Auto-Encoder[46] (AE) architecture may be an adaptive tool to cover feature extraction, generative algorithm, and information condenser. As an unsupervised model, AE allows the precision of reconstruction guaranteed by its symmetric form of the net while condensing the number of features to a minimum number.

The data to support the framework can be collected from experimental resources or simulated from various nanostructures. Experimentally, microscope-based images, ultrafast spectroscopy-based data, and other data may be used.

Finally, the extracted interatomic genome can be extended to a further generator based on the idea of Variational AE (VAE)[47]. Such extension gives the model capability of "imagination" or "induction" which may potentially cover a much broader range of interatomic cohesion.

## B. The Realization of Interatomic Genome framework: 8-value-genome AE

In this work, the interatomic genome framework is realized in a brief way and challenged by a nanocluster gene abstraction task. The classification in TP-AuNCs is the one with the fewest bond classification reported in the field of interatomic investigation which leave the interatomic genome approach a tough puzzle to resolve.

As shown in **Figure 2a->b->d->c**, targeted to the topology shift in interatomic cohesions, the interatomic genome framework in this realization works as follows: collect the geometry data and perform DFT simulations, generate ELF images and their derivatives for bosonic-fermionic description, feed the images to the AE, and then perform extensive optimization and information condensing. The AE finally yields the condensed features as interatomic genome with their statistics, an interatomic gene encoder (IGE) and a genome reader which allows reconstruction of further ELF images.

At the data collecting phase, considering the quantum facts, ELF images and their derivatives are selected but all geometrical descriptions (including bond length, bond angle, dihedral, etc.) of chemical bonds are discarded. Each atomic cohesion in a TP-AuNC in the dataset is converted to a 200×200 pixels image with two atoms placed at the leftmost center and rightmost center. Various types of S-Au and Au-Au interactions are covered and represented by images, resulting in 26,528 interatomic cohesion images. Among these, 8,454 correspond to S-Au bonds, and 18,074 correspond to Au-Au bonds.

The IGE is constructed based on a convolutional neural network (CNN) to extract features in images of interatomic cohesion. And these features are regarded as interatomic genes. So, the IGE establish a correspondence between interatomic cohesions and their genomes. The genome reader can function independently as a VAE, allowing the generation of atomic cohesion images using artificially designed genomes. This greatly broadens the research scope of interatomic genomes.

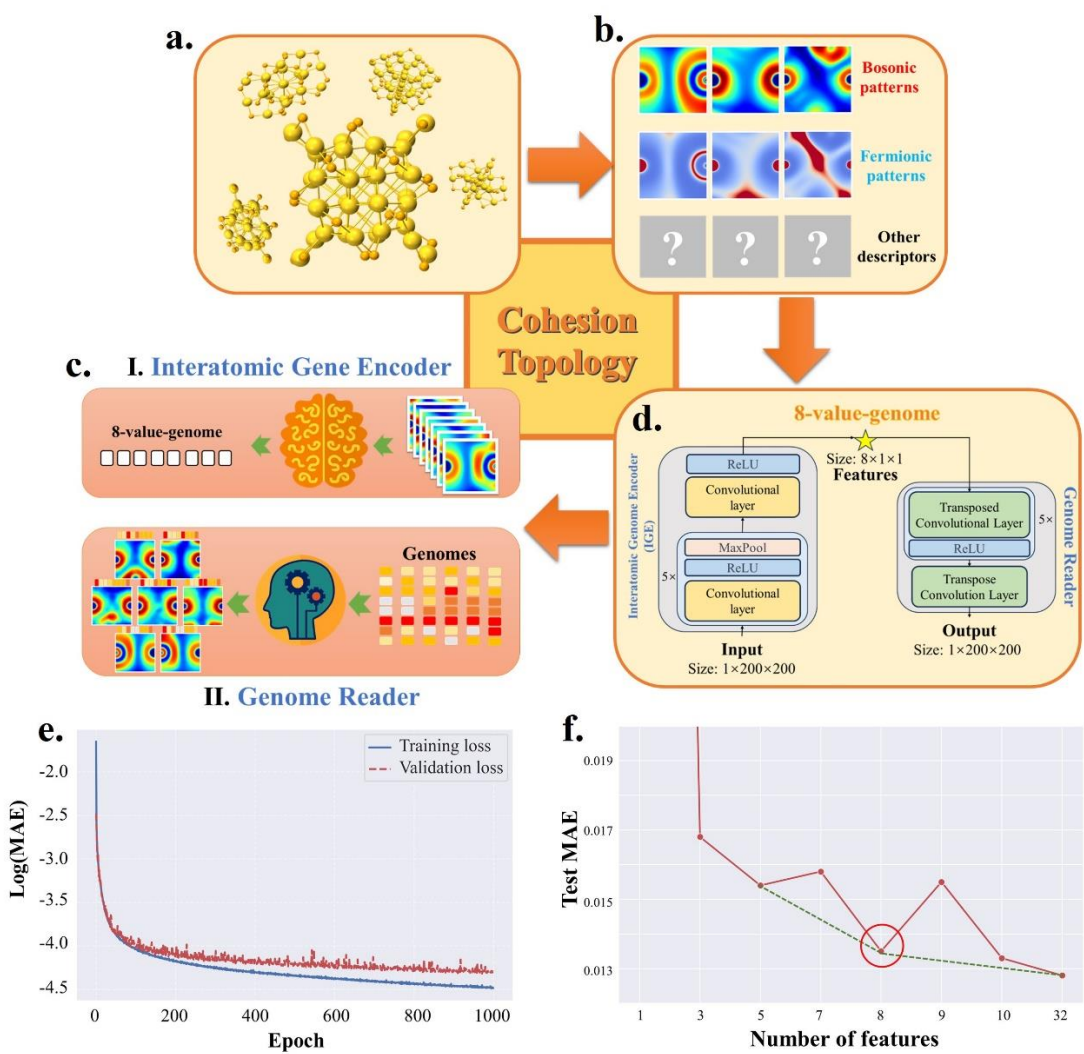

**Figure 2.** (a) Geometry structures of some TP-AuNCs in dataset. (b) ELF images and their derivatives of interatomic cohesions obtained through DFT simulations. (c) Interatomic cohesions could be deciphered as 8-value-genomes by the interatomic gene encoder. Correspondingly, 8-value-genomes consisting of numbers in specific ranges could be used to generate interatomic cohesion images by the genome reader. (d) The network architecture of the 8-value-genome AE. (e) The logarithm of MAE as a function of epochs in training and validation progress of the 8-value-genome AE. **(f)** The pixel-level MAE of reconstructing images on the test set for models extracting different lengths of genomes. Such plot explains the reason of our pick, the 8-value-geome realization.

### IV. The length of interatomic genome is just 8?

To our surprise, the answer given by the 8-value-genome AE is: even in the field of nanocluster research, the interatomic genome merely contains 8 values. After thousands of training epochs and dozens of parameters tuning, the IGE and the genome reader compromise with each other and agree to keep 8 values to trade for a mean absolute error (MAE) as low as 1.35% per pixel in reconstructing images after 1000 epochs of training **Figure 2e**. The low MAE in this work indicates the high fidelity of interatomic cohesion.

The 8-value-length interatomic genome is discovered based on extensive explorations on the network structure, the length of the interatomic genome ranges from 1 to 32 and optimizing the reconstruct qualification. As shown in **Figure 2f**, when the length of the genome is increased to 8, a significant reduction in reconstruction error can be observed. Besides, further increases in the length of the genome show limited improvement in reconstructing precision. In addition, a more than 8-value-genome usually produces spare gene(s) as shown in **Figure S3** (9- or 10-value-genome leads to 1 or 2 empty genes). Therefore, the pick of length in the interatomic genome represented the well-known "elbow" strategy. And we believe that 8-value-length should be the style of the genome.

To understand what the genes represent, the distributions of every gene are shown in **Figure 3a**. The role of each gene is further investigated based on their distributions and human interpretation via the control variable method: genome with all gene values fixed except the one or two being tested, are input into the generative model for an image to read.

#### A. The backbone of a bond: the 3rd and 8th genes

The "backbone" of each chemical bond is constituted of genes 3 and 8 (the left panel in **Figure 3b**): the 3rd gene provides the common pattern while the 8th gene trims and corrects the shape. As shown in **Figure 3a**, there are distinct differences in gene value distributions between S-Au and Au-Au bonds in the 3rd and 8th genes. As depicted in the left panel in **Figure 3c**, the elevation of the 3rd gene value results in an intensified ELF function (the increase of $I_{b-f}$, see **Equation S2**) with a general appearance resembling that of an S-Au bond, but there are still significant errors in details. Besides, the 8th gene corrects the elongated shape and smooths the bosonic/fermionic regions with the reduction of bosonic-fermionic contrast (the left panel in **Figure 3b**). Eventually, the inclusion of the 8th gene rectified most of the detailed errors present in the rough S-Au bond ELF image generated solely from the 3rd gene.

#### B. Transformation from one bond to another: the 4th and 5th genes

The 4th and 5th genes serve as switches for the interconversion from S-Au bonds to Au-Au bonds with unique Au-Au interatomic gene encoded. Even though the values of the 4th and 5th genes can be insignificant compared with other genes, a S-Au bond immediately turns into an Au-Au bond. Furthermore, the 4th gene forges the bosonic behavior in the outer shell in a pronounced and contracted manner which corresponds to the recovery of an Au atom to an isolated state (see the second Au atom in the left column in **Figure 3d**). The 5th gene introduces extra Au atom(s) for their intrusion impacts on the bosonic characteristics (see the second Au atoms in the middle column in **Figure 3d**). Besides, the fermionic features controlled by the 4th and 5th genes can be identified on the saddle points of images (located at the centers of ELF images) in **Figure 3d**. As both features are turned on, weak bosonic features with small ELF values appear in a broad interatomic region (see the second Au atoms in the right column in **Figure 3d**).

#### C. Auxiliary genes for hybridization: the 1st, 2nd, 6th and 7th genes

To compromise a variety of bindings, hybridization in nanoclusters is encoded as genes and combinations of genes. With the backbones set to the averaged values, the 1st and 6th genes describe the hybridization of the S atoms. When the 1st gene is set to 40 (see the bottom image in the column for gene 1 in **Figure 3e**), the bosonic regions appear to be the three vertices of a $sp^2$-hybridized triangular configuration around the S atom. The 1st and 6th genes emphasize different orientations which allow the capability to describe bonding from different sites. Hybridization can also be achieved by multiple genes. As shown in **Figure 4c**, if both the 1st and 6th genes are added, it will result in $sp^3d$ hybridization of the S atom. Upon opening the 4th and 5th genes, adding both the 2nd and 7th genes will lead to $sp^3d^2$ hybridization of the Au atom. By collectively adjusting the values of the 1st, 2nd, 6th, and 7th genes, various complex hybridizations can be generated. The 4 auxiliary genes, may give enough variations of hybridizations to nanoclusters even with the ones that cannot be classified in a straightforward way.

Besides, with **Equation S1**, independently adjusting each of these 4 genes alters the bosonic-fermionic tendencies of electrons in different regions. For example, in the case of the 2nd and 7th genes (see the column for genes 2 and 7 in **Figure 3e**), an increase in the value of the 2nd gene gradually leads to the behavior of electrons within the entire low ELF value region becoming more bosonic. An elevation in the 7th gene significantly enhances the fermionic nature of localized electrons within the low ELF value region. The 1st and 6th genes also display regional impact on the fermionic characteristics.

Based on the interatomic genome with associated distributions collected, one can resample genes and substitute them into the genome reader to explore the "imaginations" of the generative model demonstrated in **Figure 1**. Further investigations focus on how the interatomic genome is descriptive.

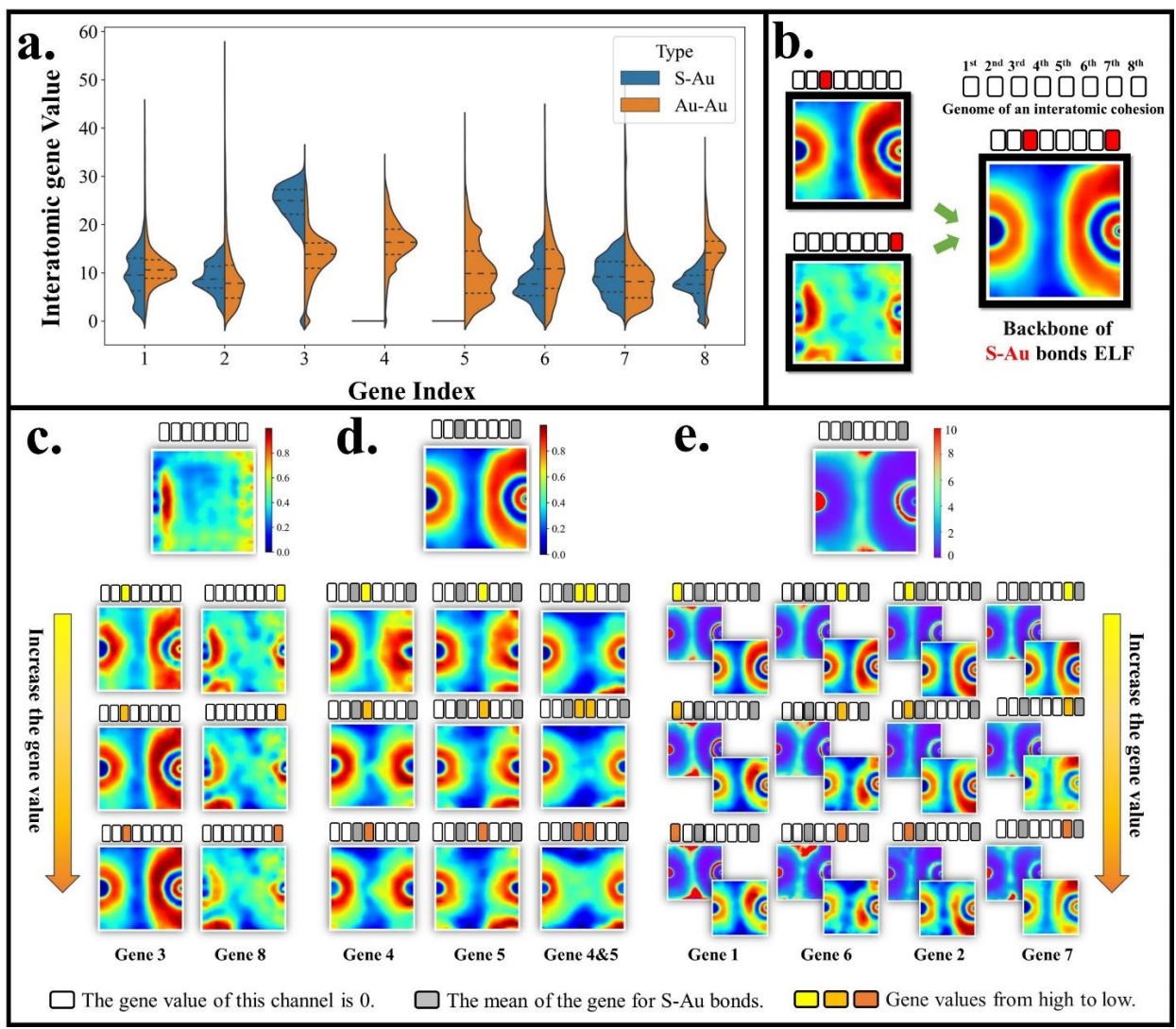

**Figure 3. (a)** Violin plots depicting the values distribution of each gene extracted by the IGE for S-Au bonds and Au-Au bonds in the dataset. In the violin plot, wider sections indicate that the distribution of values within that range in the gene is denser. **(b)** The 3rd and 8th genes collectively form the backbone of the interatomic cohesion. When the 4th and 5th genes, acting as switches for Au-Au bonds, are closed, the 3rd and 8th genes constitute the backbone for S-Au bonds; when the switches are opened, they form the backbone for Au-Au bonds. **(c), (d), and (e)** Utilizing the method of controlling variables, independently elevate the value of the gene under test. ELF image in **(c), (d), and (e)** are generated by inputting test genomes into the genome reader. The generated images undergo transformation via Equation S1 to represent changes in the fermionic component $\chi_\sigma$ within interatomic cohesions.

## V. Machines do not interpret bonds as human did

In addition to the gene-wised interpretation, one can let the genes talk by reading their associated interatomic cohesion out of the "imaged scenarios" obtained from the genome reader. The interatomic genome contains no bond lengths, bond angles, or dihedrals directly as human's focus, but they do cover enormous situations in molecules or even in crystals, much more than the data set contains. Let us do a simple math: if each gene is simplified to a set of 6 discrete values, the 8-value-genome piece gives about 1.7 million ($6^8$=1.7 million) interatomic cohesions. The 8-value-genome, in its own way, supported a wide coverage of the interatomic cohesions, beyond the training data it saw. In the following description, we show some generated patterns (called machine's imaginations) that can be identified to our knowledge. These imaginations may not be existing ones but extensions of some inner patterns that are controllable by the values of genes.

### A. Description in chemistry style: polarization and hybridization

Machine's imaginations include some chemistry-style approaches. As shown in **Figure 4a-c**, a large portion of imaginations resembles classical chemical bond theories and displaces numerous bonding configurations. By manipulating 1 gene value, the corresponding interatomic cohesion shifts from strong polarizations to much weaker ones (**Figure 4a**). In addition, hybridization can be similarly adjusted by changing another gene value. The top row in **Figure 4b** demonstrated the change from triangular bipyramidal to heptagonal bipyramidal shape. The bottom row in **Figure 4b** shows the change from the SP to SP$^3$ hybridization on an S atom. Besides, the replications of typical hybridizations can also be observed as demonstrated in **Figure 4c**.

### B. Description in solid state physics: atom alignments and crystal orientation

Additionally, the interatomic genome does not limit itself to molecules but also extend to scenarios in solids. As shown in **Figure 4d**, one can see interatomic patterns in a face-centered cubic (FCC) lattice along different orientations which is controlled by the 7th gene value. As shown in **Figure 4f**, the associated images represent the gradual shift in the alignment of gold atoms from a parallel one to a non-parallel one. Such alignment cannot be expected in a real crystal of gold, but it belongs to the extension in the imaginations of the generative model.

### C. Description in impacted geometry: atomic motions

The 8-value-genome can also capture cases where interatomic cohesions are impacted by some other atoms. **Figure 4e** illustrates the bending of the Au-S-Au motif, a commonly seen pattern in TP-AuNCs. Since motifs serve as mounting interfaces of metal cores and ligands, it is "rational" for interatomic genome to capture the bending.

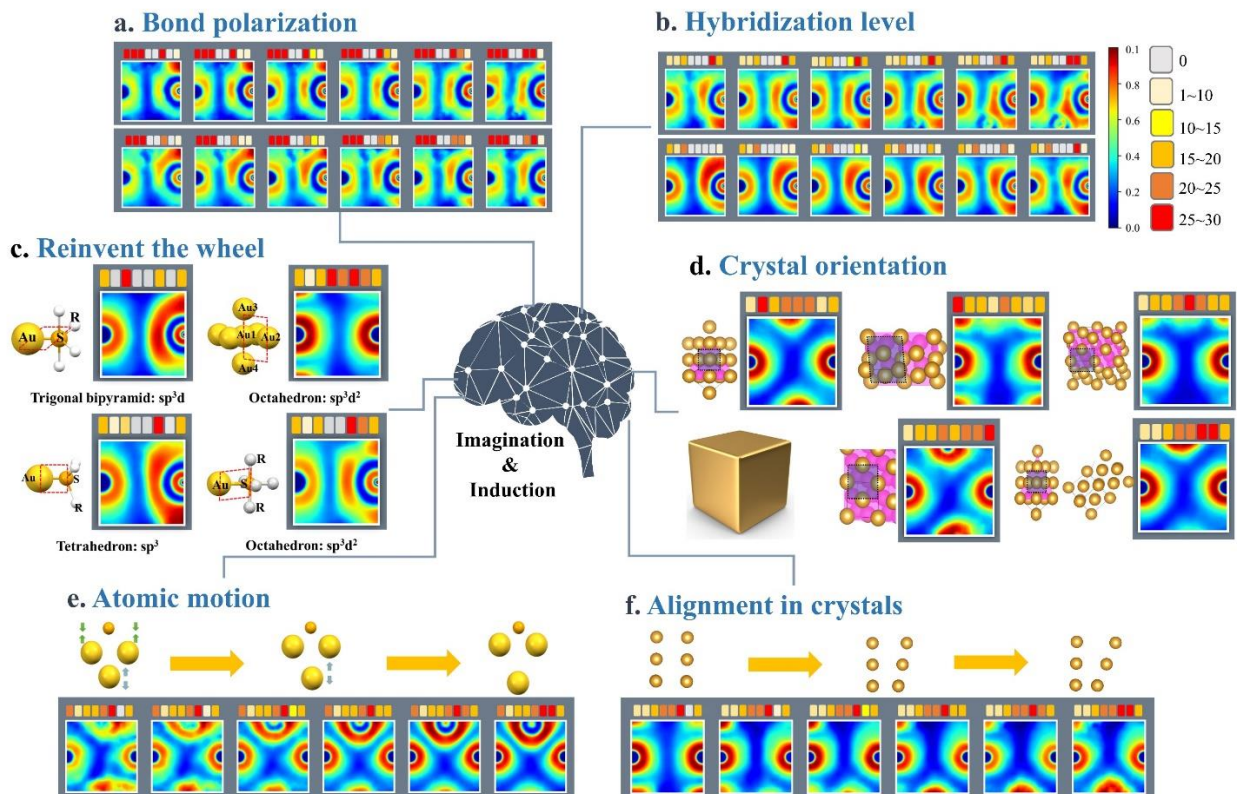

**Figure 4.** **(a)** and **(b)** The machine's imagination regarding evolutions of bond polarization and hybridization level. These images are generated based on given genomes through continuous adjustments of a single gene value. In other words, only the initial state of the bond is provided, and subsequent changes are imagined by the algorithm based on its own bosonic/fermionic quantitative standards. **(c)** The correspondence between the generated images and the hybridization in TP-AuNCs. **(d)** The machine's imagination of Au-Au bonds in crystals with different crystal orientation. **(e) and (f)** The machine's imagination regarding evolutions of atomic motion and alignment.

### VI. A Roadmap and Applications

From the whole dataset, the similarity between two arbitrary interatomic cohesions can be evaluated as the statistical correlation between their 8-value-genome. Thus, one can consider all captured interatomic cohesion as nodes with edges that connect similar ones (correlation > 0.9) for a complex network as the evolutionary roadmap of interatomic cohesions (**Figure 5a**). The colors represent the PageRank scores of nodes and the highest score in each zone is plotted as the most important one in the neighborhood of genome similarity. S-Au interactions are concentrated to a small island which reflects the homogeneity among such interactions. The Au-S-Au motifs are located on the two upper islands. The mainland is constituted of "peaks" of 6 Au-Au patterns: two with strong interatomic fermionic patterns, two with weak interatomic fermionic patterns, and two with impacts from extra gold atom(s). Such an evolutionary roadmap forms the support for the generative model to develop further rational imaginations.

As some chemical/physical patterns are impacted, one can upgrade the interatomic genome to a dynamic one to apply them to various interatomic cohesion-related tasks. Our demonstration of the IGE's application in 3 different research topics including the chemisorption, molecular dynamic, and attosecond optical process. **Figure 5b** demonstrates the interatomic genome changes as the approach of an $H_2O_2$ molecule towards the $Au_{12}$ cluster as a common catalytic scenario. To evaluate the adhesion process, 4 consecutive configurations along the adhesion are analyzed which gives an adhesion dynamic interatomic genome ($H_2O_2$ approaching vs. 8-value-genome). As indicated by the 3rd gene, the backbone of the focused S-Au interaction is weakened while the hybridization of it is enhanced. Such analysis allows us to further process the interatomic cohesion or feed the genes to other machine-learning models to find the patterns behind them.

In the molecular dynamical application (**Figure 5c**), an $Au_{12}$ cluster with 8 glutathione ligands is simulated to represent the flexibility of a cluster's ligand. Due to the solubility and hydrogen bonds in glutathione, the geometry shape is complicated with typical bond geometry. Here as represented by the dynamical plot, a selected Au-Au bond length shows significant changes in its interatomic genome. The backbone gene does not change much, but the polarization and hybridization genes change for a more significant change in fermionic patterns within the gold core.

Besides, the attosecond optical process is also simulated with real time Time-dependent (DFT) for the study of a gold nanocluster in the presents of attosecond laser pulses. As the dipoles response to the electron field differently, A time-dependent "kick" simulation allows us to find different electronic resonance modes with associated 8-value-genome analyzed as shown in **Figure 6a**. Considering the direction of the external pulse, an Au-Au bond is focus with all its genes deciphered. As demonstrated in **Figure 6c**, the on-resonance pulse induces the electron oscillation easily even after the external pulse vanishes. In contrast, the off-resonance pulse does not cause that significant oscillation in the cluster **Figure 6b**. As shown in **Figure 6d**, the on-resonance 4.9eV pulse mainly changes the 3rd and 8th genes which means that the 4.9eV pulse significantly alters the backbone of the Au-Au interaction. The pulse also magnified the intrusion of other gold atoms (the 5th gene) and intensified the fermionic characteristics in the Au-Au interaction (the 7th gene).

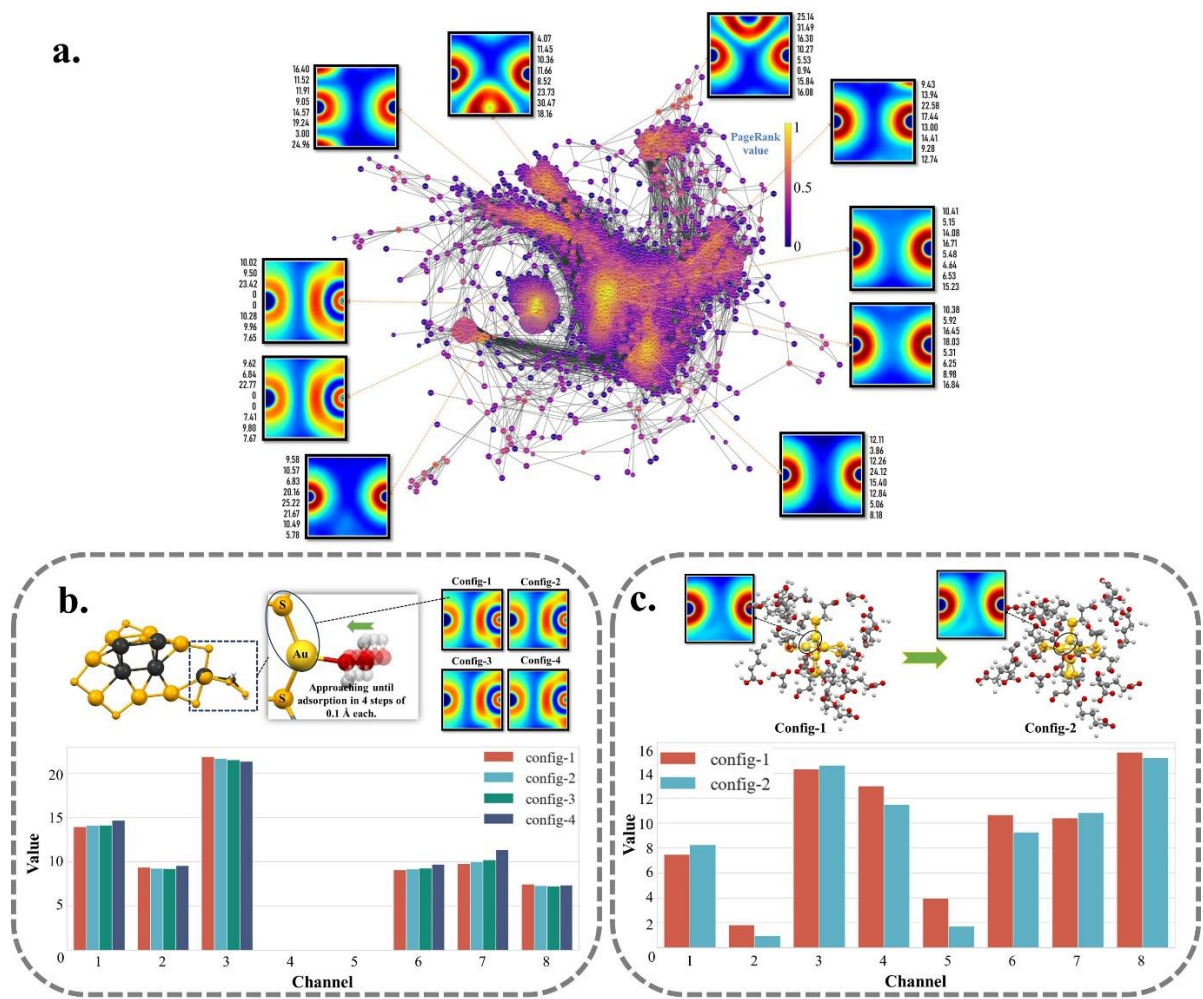

**Figure 5.** **(a)** The evolutionary roadmap of interatomic cohesions: each node represents an image; each edge connects two nodes with a correlation (similarity) greater than 0.9. S-Au interactions are located on the "island" in the middle surrounded by Au-Au interactions. Colors of the network are obtained from the PageRank value of each vertex. **(b)** An application of the interatomic genome on adsorption of $H_2O_2$ on an $Au_{12}$ NC. The adsorption process is divided into four equally spaced approaches of $H_2O_2$, each approaching the adsorption site by 0.01 Å. Each critical step is plotted by its kinetic process with geometry change from the yellow

ball-stick to the dark ball-stick structures. **(c)** An application of the interatomic genome for ligand dynamic task: a glutathione-protected $Au_{12}$ NC with its ligand oscillates in a complicated pattern in water.

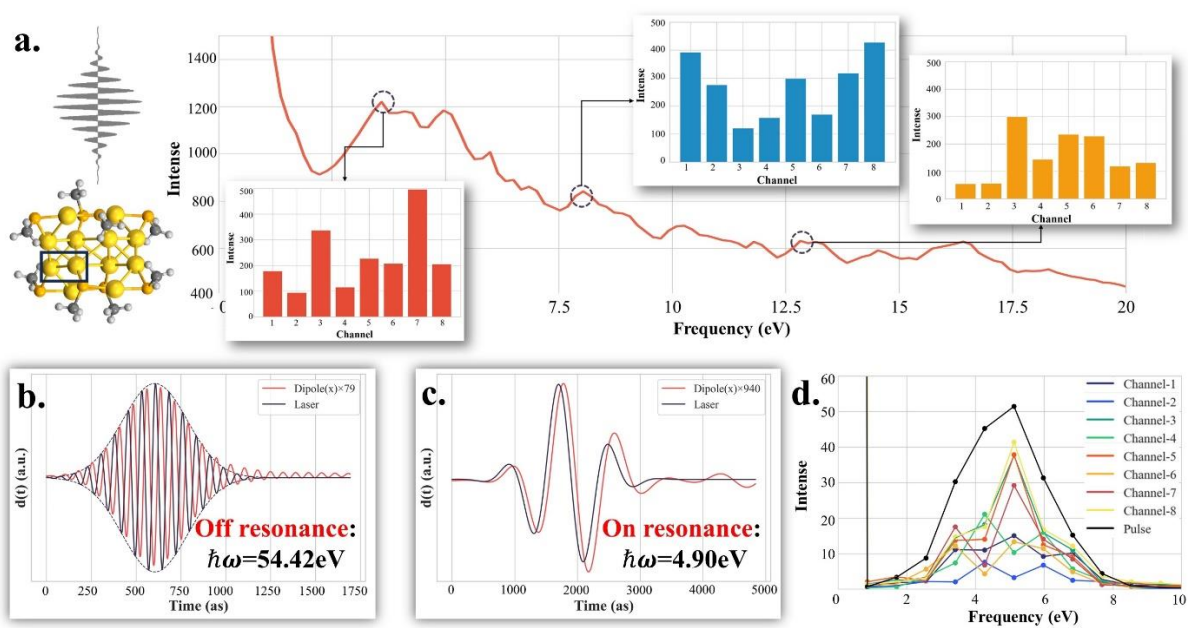

**Figure 6.** An application of the interatomic genome on the ultrafast (atto- or femto-second) processes: **(a)** absorption spectrum in a gold cluster; **(b)** an off-resonance pulses with induced dipole in the cluster; **(c)** on-resonance pulse with induces dipoles in the cluster; **(d)** the corresponding spectrum of pulse vs induced changes in the 8-value-genome.

## VII. SUMMARY

In summary, this work introduced a new data-driven framework combining deep learning algorithms, molecular/crystal geometry, electron structure simulation, bosonic-fermionic characteristics, and information condenser with an encoder for reading and a generative model for producing its result. In our realization reported, the ELF images and its derivatives, AE and Variation AE are considered which allows the generation and condensing of interatomic features to an 8-value-genome. The reconstruction precision guarantees succinct encapsulation of full interatomic information in the 8-value-genome.

The framework focuses on the bosonic-fermionic topology in great detail. The bottleneck in the lack of training data relies on the idea of discarding the geometry of bonds and capturing merely interatomic patterns with the current layout (putting 2 atoms at left and right centers). Fortunately, our training results and interpretation approaches support such a resolution. Furthermore, our framework unleashes the power of the neural network in interatomic patterns which surpasses human knowledge accumulated in the last 100 years. Yet our model is a coherent model instead of a model with an enormous collection of superficial statistics. The interatomic 8-value-genome speaks for itself. The value-wise, pattern-wise, and dataset-wise analysis bring further bond-related tasks to a new level. Such a framework can be extended and applied to materials or systems with equivalent or even more complex structures to express their electronic structures and interaction information in a feature-based quantitative style. With our framework, we may want to seek another answer: how many values does the interatomic genome contain to cover the interactions across the whole periodic table?

## ASSOCIATED CONTENT

Please see the supplementary information.

## AUTHOR INFORMATION

## Corresponding Author

Yonghui Li, Email: yonghui.li@tju.edu.cn;Yonghui Li, Email: yonghui.li@tju.edu.cn;

## AUTHOR CONTRIBUTIONS

Xinxu Zhang designed and trained the model. Yonghui Li conceptualization the model, analysis algorithm and visualizations. Jiahao Wei performed simulations for gold nanocluster under attosecond pulses. Xinxu Zhang, Hui Jia and Jiamin Liu collected the cluster data in the dataset and performed the DFT calculations required for constructing the dataset. Xinxu Zhang, Guo Li, Yulong Wu and Ling Liu organized and analyzed the test data. Yonghui Li and Changlong Liu reviewed the results. Xinxu Zhang and Yonghui Li prepared the manuscript. All authors discussed the results and commented the manuscript.

## ACKNOWLEDGMENT

We thank Carsten A. Ullrich from University of Missouri for valuable comments and suggestions. Yonghui Li was sponsored by the National Key Research and Development Program of China (Grant No.2021YFF1200701), the National Natural Science Foundation of China (Grant No.11804248) and the Key Projects of Tianjin Natural Fund 21JCZDJC00490. Changlong Liu was supported by the National Natural Science Foundation of China (No. 11675120 and 11535008) and the National Key Research and Development Program of China (No. 2021YFF1200700). Xiaodong Zhang was supported by the National Key Research and Development Program of China (No. 2021YFF1200700), the National Natural Science Foundation of China (Grant Nos. 91859101, 81971744, U1932107, and 82001952).

## REFERENCES

(1) Gillespie, R. J.; Robinson, E. A. Gilbert N. Lewis and the Chemical Bond: The Electron Pair and the Octet Rule from 1916 to the Present Day. *Journal of computational chemistry* **2007**, *28* (1), 87–97.

(2) Bader, R. F.; Gillespie, R. J.; MacDougall, P. J. A Physical Basis for the VSEPR Model of Molecular Geometry. *Journal of the American Chemical Society* **1988**, *110* (22), 7329–7336.

(3) Gillespie, R. J.; Hargittai, I. *The VSEPR Model of Molecular Geometry*; Courier Corporation, 2013.

(4) Mu, X.; Wang, J.; Li, Y.; Xu, F.; Long, W.; Ouyang, L.; Liu, H.; Jing, Y.; Wang, J.; Dai, H.; Liu, Q.; Sun, Y.; Liu, C.; Zhang, X.-D. Redox Trimetallic Nanozyme with Neutral Environment Preference for Brain Injury. *ACS Nano* **2019**, *13* (2), 1870–1884. https://doi.org/10.1021/acsnano.8b08045.

(5) Vibrational Spectral Studies, Quantum Mechanical Properties, and Biological Activity Prediction and Inclusion Molecular Self-Assembly Formation of N-N'-Dimethylethylene Urea. *Biointerface Res Appl Chem* **2021**, *12* (3), 3996–4017. https://doi.org/10.33263/BRIAC123.39964017.

(6) Abu Ali, O. A.; Elangovan, N.; Mahmoud, S. F.; El-Bahy, S. M.; El-Bahy, Z. M.; Thomas, R. Synthesis, Structural Features, Excited State Properties, Flouresence Spectra, and Quantum Chemical Modeling of (E)-2-Hydroxy-5-(((4-Sulfamoylphenyl)Imino) Methyl)Benzoic Acid. *Journal of Molecular Liquids* **2022**, *360*, 119557. https://doi.org/10.1016/j.molliq.2022.119557.

(7) Fatima, A.; Khanum, G.; Savita, S.; Pooja, K.; Verma, I.; Siddiqui, N.; Javed, S. Quantum Computational, Spectroscopic, Hirshfeld Surface, Electronic State and Molecular Docking Studies on Sulfanilic Acid: An Anti-Bacterial Drug. *Journal of Molecular Liquids* **2022**, *346*, 117150. https://doi.org/10.1016/j.molliq.2021.117150.

(8) Guo, Q.; Wu, Y.; Xia, L.; Yu, X.-F.; Zhang, K.; Du, Y.; Zhang, L.; Tang, H.; Cheng, J.; Shang, J.; Peng, Y.; Li, Z.; Man, X.; Yang, X. Stitching Electron Localized Heptazine Units with "Carbon Patches" to Regulate Exciton Dissociation Behavior of Carbon Nitride for Photocatalytic Elimination of Petroleum Hydrocarbons. *Chemical Engineering Journal* **2023**, *452*, 139092. https://doi.org/10.1016/j.cej.2022.139092.

(9) Becke, A. D.; Edgecombe, K. E. A Simple Measure of Electron Localization in Atomic and Molecular Systems. *The Journal of Chemical Physics* **1990**, *92* (9), 5397–5403. https://doi.org/10.1063/1.458517.

(10) Savin, A.; Becke, A.; Flad, J.; Nesper, R.; Preuss, H.; Von Schnering, H. A New Look at Electron Localization. *Angewandte Chemie International Edition in English* **1991**, *30* (4), 409–412.

(11) Kohout, M.; Wagner, F. R.; Grin, Y. Electron Localization Function for Transition-Metal Compounds. *Theoretical Chemistry Accounts* **2002**, *108*, 150–156.

(12) Silvi, B.; Savin, A. Classification of Chemical Bonds Based on Topological Analysis of Electron Localization Functions. *Nature* **1994**, *371* (6499), 683–686. https://doi.org/10.1038/371683a0.

(13) Noury, S.; Krokidis, X.; Fuster, F.; Silvi, B. Computational Tools for the Electron Localization Function Topological Analysis. *Computers & chemistry* **1999**, *23* (6), 597–604.

(14) Heaven, M. W.; Dass, A.; White, P. S.; Holt, K. M.; Murray, R. W. Crystal Structure of the Gold Nanoparticle [N(C8H17)4][Au25(SCH2CH2Ph)18]. *Journal of the American Chemical Society* **2008**, *130* (12), 3754–3755. https://doi.org/10.1021/ja800561b.

(15) Jadzinsky, P. D.; Calero, G.; Ackerson, C. J.; Bushnell, D. A.; Kornberg, R. D. Structure of a Thiol Monolayer-Protected Gold Nanoparticle at 1.1 A Resolution. *Science (New York, N.Y.)* **2007**, *318* (5849), 430–433. https://doi.org/10.1126/science.1148624.

(16) Jin, R.; Li, G.; Sharma, S.; Li, Y.; Du, X. Toward Active-Site Tailoring in Heterogeneous Catalysis by Atomically Precise Metal Nanoclusters with Crystallographic Structures. *Chem. Rev.* **2021**, *121* (2), 567–648. https://doi.org/10.1021/acs.chemrev.0c00495.

(17) Shaik, S. S.; Hiberty, P. C. *A Chemist's Guide to Valence Bond Theory*; John Wiley & Sons, 2007.

(18) Weinhold, F.; Landis, C.; Glendening, E. What Is NBO Analysis and How Is It Useful? *International reviews in physical chemistry* **2016**, *35* (3), 399–440.

(19) Glendening, E. D.; Landis, C. R.; Weinhold, F. NBO 7.0: New Vistas in Localized and Delocalized Chemical Bonding Theory. *Journal of computational chemistry* **2019**, *40* (25), 2234–2241.

(20) Weinhold, F. Natural Bond Orbital Analysis: A Critical Overview of Relationships to Alternative Bonding Perspectives. *Journal of computational chemistry* **2012**, *33* (30), 2363–2379.

(21) Glendening, E. D.; Weinhold, F. Natural Resonance Theory: I. General Formalism. *Journal of computational chemistry* **1998**, *19* (6), 593–609.

(22) Glendening, E. D.; Weinhold, F. Natural Resonance Theory: II. Natural Bond Order and Valency. *Journal of Computational Chemistry* **1998**, *19* (6), 610–627.

(23) Glendening, E. D.; Badenhoop, J.; Weinhold, F. Natural Resonance Theory: III. Chemical Applications. *Journal of computational chemistry* **1998**, *19* (6), 628–646.

(24) Zubarev, D. Y.; Boldyrev, A. I. " Developing Paradigms of Chemical Bonding: Adaptive Natural Density Partitioning. *Physical chemistry chemical physics* **2008**, *10* (34), 5207–5217.

(25) Zhao, L.; Pan, S.; Holzmann, N.; Schwerdtfeger, P.; Frenking, G. Chemical Bonding and Bonding Models of Main-Group Compounds. *Chem. Rev.* **2019**, *119* (14), 8781–8845. https://doi.org/10.1021/acs.chemrev.8b00722.

(26) Maseras, F.; Morokuma, K. Application of the Natural Population Analysis to Transition-Metal Complexes. Should the Empty Metal p Orbitals Be Included in the Valence Space? *Chemical physics letters* **1992**, *195* (5–6), 500–504.

(27) Bayse, C. A.; Hall, M. B. Prediction of the Geometries of Simple Transition Metal Polyhydride Complexes by Symmetry Analysis. *Journal of the American Chemical Society* **1999**, *121* (6), 1348–1358.

(28) Diefenbach, A.; Bickelhaupt, F. M.; Frenking, G. The Nature of the Transition Metal- Carbonyl Bond and the Question about the Valence Orbitals of Transition Metals. A Bond-Energy Decomposition Analysis of TM (CO) 6q (TMq= Hf2-, Ta-, W, Re+, Os2+, Ir3+). *Journal of the American Chemical Society* **2000**, *122* (27), 6449–6458.

(29) Bader, R. F.; Stephens, M. E. Spatial Localization of the Electronic Pair and Number Distributions in Molecules. *Journal of the American Chemical Society* **1975**, *97* (26), 7391–7399.

(30) Bader, R. F.; Beddall, P. Virial Field Relationship for Molecular Charge Distributions and the Spatial Partitioning of Molecular Properties. *The Journal of Chemical Physics* **1972**, *56* (7), 3320–3329.

(31) Cremer, D.; Kraka, E. Chemical Bonds without Bonding Electron Density—Does the Difference Electron-Density Analysis Suffice for a Description of the Chemical Bond? *Angewandte Chemie International Edition in English* **1984**, *23* (8), 627–628.

(32) Mitoraj, M. P.; Michalak, A.; Ziegler, T. A Combined Charge and Energy Decomposition Scheme for Bond Analysis. *Journal of chemical theory and computation* **2009**, *5* (4), 962–975.

(33) Michalak, A.; Mitoraj, M.; Ziegler, T. Bond Orbitals from Chemical Valence Theory. *The Journal of Physical Chemistry A* **2008**, *112* (9), 1933–1939.

(34) Zhang, Q.; Li, W.-L.; Xu, C.-Q.; Chen, M.; Zhou, M.; Li, J.; Andrada, D. M.; Frenking, G. Formation and Characterization of the Boron Dicarbonyl Complex [B (CO) 2]-. *Angewandte Chemie International Edition* **2015**, *54* (38), 11078–11083.

(35) Andrada, D. M.; Frenking, G. Stabilization of Heterodiatomic SiC through Ligand Donation: Theoretical Investigation of SiC (L) 2 (L= NHCMe, CAACMe, PMe3). *Angewandte Chemie International Edition* **2015**, *54* (42), 12319–12324.

(36) Bollermann, T.; Cadenbach, T.; Gemel, C.; von Hopffgarten, M.; Frenking, G.; Fischer, R. A. Molecular Alloys: Experimental and Theoretical Investigations on the Substitution of Zinc by Cadmium and Mercury in the Homologous Series [Mo (M′ R) 12] and [M (M′ R) 8](M= Pd, Pt; M′= Zn, Cd, Hg). *Chemistry–A European Journal* **2010**, *16* (45), 13372–13384.

(37) von Hopffgarten, M.; Frenking, G. Building a Bridge between Coordination Compounds and Clusters: Bonding Analysis of the Icosahedral Molecules [M (ER) 12](M= Cr, Mo, W; E= Zn, Cd, Hg). *The Journal of Physical Chemistry A* **2011**, *115* (45), 12758–12768.

(38) Norsko, J. Chemisorption on Metal Surfaces. *Reports on Progress in Physics* **1990**, *53* (10), 1253.

(39) Nørskov, J. Electronic Factors in Catalysis. *Progress in surface science* **1991**, *38* (2), 103–144.

(40) Zhang, Q.; Guo, L. Mechanism of the Reverse Water–Gas Shift Reaction Catalyzed by Cu 12 TM Bimetallic Nanocluster: A Density Functional Theory Study. *Journal of Cluster Science* **2018**, *29*, 867–877.

(41) Gorzkowski, M. T.; Lewera, A. Probing the Limits of D-Band Center Theory: Electronic and Electrocatalytic Properties of Pd-Shell–Pt-Core Nanoparticles. *The Journal of Physical Chemistry C* **2015**, *119* (32), 18389–18395.

(42) Sun, Z.; Song, Z.; Yin, W.-J. Going beyond the D-Band Center to Describe CO2 Activation on Single-Atom Alloys. *Advanced Energy and Sustainability Research* **2022**, *3* (2), 2100152.

(43) Zhong, S.; Hu, J.; Yu, X.; Zhang, H. Molecular Image-Convolutional Neural Network (CNN) Assisted QSAR Models for Predicting Contaminant Reactivity toward OH Radicals: Transfer Learning, Data Augmentation and Model Interpretation. *Chemical Engineering Journal* **2021**, *408*, 127998.

(44) Omidvar, N.; Pillai, H. S.; Wang, S.-H.; Mou, T.; Wang, S.; Athawale, A.; Achenie, L. E.; Xin, H. Interpretable Machine Learning of Chemical Bonding at Solid Surfaces. *The Journal of Physical Chemistry Letters* **2021**, *12* (46), 11476–11487.

(45) Perdew, J. P.; Burke, K.; Wang, Y. Generalized Gradient Approximation for the Exchange-Correlation Hole of a Many-Electron System. *Physical review B* **1996**, *54* (23), 16533.

(46) Wang, Y.; Yao, H.; Zhao, S. Auto-Encoder Based Dimensionality Reduction. *Neurocomputing* **2016**, *184*, 232–242.

(47) Kingma, D. P.; Welling, M.; others. An Introduction to Variational Autoencoders. *Foundations and Trends® in Machine Learning* **2019**, *12* (4), 307–392.

----------------------------